\documentclass[preprint,showpacs,preprintnumbers,amsmath,amssymb]{revtex4}

\usepackage{graphicx}% Include figure files
\usepackage{dcolumn}% Align table columns on decimal point
\usepackage{bm}% bold math

\begin{document}

\preprint{APS/123-QED}

%\title{An Information Theory Approach for the Construction of a Similarity Index for Atoms}
\title{A Quantum Similarity Study of Atomic Density Functions: Insights from Information Theory and the Role of Relativistic Effects}

\author{A.~Borgoo$^{+}$}
\author{M.~Godefroid$^{++}$}
\author{P.~Indelicato$^{+++}$}
\author{F.~De~Proft$^{+}$}
\author{P.~Geerlings$^{+*}$}

\affiliation{$^{+}$Vrije Universiteit Brussel (Free University of Brussels VUB), 
Departement of General Chemistry (ALGC),
Pleinlaan, 2, 
1050 Brussels Belgium}

 \affiliation{$^{++}$Service de Chimie quantique et Photophysique - CP160/09, 
Université Libre de Bruxelles, 
Av. F.D. Roosevelt 50, 
1050 Brussels Belgium}

\affiliation{$^{+++}$Laboratoire Kastler Brossel, École Normale Supérieure (ENS) et
 Université Pierre et Marie Curie-Paris 6, F-75231 Paris CEDEX 05, France}

%corauth[cor1]{Tel.: +32~2~6293314, 
%Fax: +32~2~6293317 \ead{pgeerlin@vub.ac.be}}

\begin{abstract}
A novel Quantum Similarity Measure (QSM) is constructed, based on concepts from Information Theory. In an application of QSM to atoms, the new QSM  and its corresponding Quantum Similarity Index (QSI) are evaluated throughout the periodic table, using the atomic electron densities and shape functions calculated in the Hartree-Fock approximation. The periodicity of Mendeleev's Table is regained for the first time through the evaluation of a QSM. Evaluation of the Information Theory based QSI demonstrates however that the patterns of periodicity are lost due to the renormalization of the QSM, yielding chemically less appealing results for the QSI. A comparison of the information content of a given atom on top of a group with the information content of the elements in the subsequent rows reveals another periodicity pattern.

Relativistic effects on the electronic density functions of atoms are investigated. Their importance  is quantified in a QSI study by comparing for each atom, the density functions evaluated in the Hartree-Fock and Dirac-Fock approximations. The smooth decreasing of the relevant QSI along the periodic table illustrates in a quantitative way the increase of relativistic corrections with the nuclear charge.
\end{abstract}

%Use showkeys class option if keyword
\maketitle

\section{Introduction}
Quantification of molecular similarity using an electron  density based Quantum Similarity Index (QSI) yields compact information on the similarity in shape and extent of the electron density distribution of various molecules. These data can be used as descriptor in comparative discussions of molecular structure and reactivity \cite{nieuw}. In recent years one notices a multitude of methodological studies on various aspects of Quantum Similarity of molecules such as the use of different separation operators \cite{vijf}, the replacement of the density by more appropriate, reactivity oriented functions \cite{zes,zeven}, within the context of conceptual DFT \cite{acht} and the peculiarities arising in the comparison of enantiomers \cite{negen,tien}.

Most remarkably the field of similarity of isolated atoms remained practically unexplored, with the exception of two papers by Carb\'{o} and coworkers \cite{elf,twaalf} and a third one by the present authors \cite{CPL1}. In the first paper \cite{elf} atomic self-similarity was studied, whereas the second one \cite{twaalf} contains a relatively short study on atomic and nuclear similarities, leading to the conclusion that atoms bear the highest resemblance to their neighbors in the Periodic Table. 
In  \cite{CPL1} we reported results of the Carb\'{o} index, which turns out to mask the periodicity of Mendeleev's Table, followed by results of an Information Theory based approach, where the Information Discrimination was evaluated for atomic electron density functions.  
Hereby the periodicity in the atomic electron density and shape functions throughout Mendeleev's Table was regained. 
The relevance of Information Theory in applications to electron density functions of atoms has also been confirmed in a study of the $N$-derivative of the Shannon entropy of shape functions \cite{Sen}. 
 
The present work is, in a first part, concerned with the construction of a new Quantum Similarity Measure (QSM) along the lines of the study in  \cite{CPL1}. The simplification of the density function based Quantum Similarity Index to a shape based expression emphasizes the potential of the shape function as an alternative to the density function.
The use of the shape function as a fundamental descriptor of atomic and molecular systems is indeed a current topic of investigation in the domain of Quantum Chemical description of atoms and molecules~\cite{shapefunction,vijftien,newref}. 
After defining the mentioned \emph{Information Theory based} QSM and the corresponding QSI, they are evaluated for all pairs of atoms in the Periodic Table. The results are interpreted and investigated for patterns of periodicity. 

In the second part of this work we investigate the relativistic effects on the electron density functions of atoms and their quantification using QSI. From the relativistic effects on total energies one can infer these effects have implications for the electron densities, as visualized in \cite{Sen} for the Rn atom. The effect of relativity on atomic wave functions has been studied in the pioneering work of Burke and Grant \cite{nref1} who presented graphs and tables to show the order of magnitude of corrections to the hydrogenic charge distributions for $Z=80$. The relative changes in the binding energies and expectation values of $r$ due to relativistic effects are known from the comparison of the results obtained by solving both the Schrödinger and Dirac equations for the same Coulomb potential. The contraction of the $ns$-orbitals is a well known example of these relativistic effects. But as pointed out by Desclaux in his ``{\it Tour historique}''~\cite{nref2}, for a many-electron system, the self-consistent field effects change this simple picture quite significantly. Indeed, contrary to the single electron solution of the Dirac equation showing mainly the mass variation with velocity, a Dirac-Fock calculation includes the changes in the spatial charge distribution of the electrons induced by the self-consistent field. 
The framework of QSI offers a simple way of quantifying relativistic effects on atomic electron densities via a comparison between non relativistic Hartree-Fock and relativistic Dirac-Fock electron density functions. 

\section{Methodology}

\subsection{Quantum Similarity Indices and Information Theory}

Our work is situated in the context of a mathematically rigorous theory of Quantum Similarity Measures (QSM) and Quantum Similarity Indices (QSI) as developed by Carb\'{o} \cite{vijf,vier}.
This theory encompasses quantum objects, e.g. atomic and molecular systems.
Following Carb\'{o}, we define the similarity of two atoms ($a$ and $b$) as a QSM $Z_{ab}(\Omega)$, 
\begin{eqnarray} %\label{similariteitsdefinitie}
	& Z_{ab}(\Omega)  =  \int  \rho_{a}(\mathbf{r}_{1}) 
	\, \Omega (\textbf{r}_{1},\textbf{r}_{2}) 
	\, \rho_{b}(\mathbf{r}_{2}) \; d \mathbf{r}_{1} d \mathbf{r}_{2} \label{ZABdef} 
\end{eqnarray}
where $\Omega ({\mathbf{r}_1, \mathbf{r}_2})$ is a positive definite operator. Renormalization to
\begin{eqnarray}
	& SI_{ \Omega }  =  \frac{Z_{ab}(\Omega) } {\sqrt{Z_{aa}(\Omega) }\sqrt{Z_{bb}(\Omega) }} \label{SIdef},
\end{eqnarray}
yields a QSI $SI_{\Omega}$ with values comprised between $0$ and $1$.

\noindent By choosing the operator $\Omega (\textbf{r}_1, \textbf{r}_2)$ to be the Dirac- $\delta(\textbf{r}_{1}$-$\textbf{r}_{2})$ function, expression \ref{ZABdef} reduces to an overlap integral, yielding the simplest form of the Carb\'{o} Similarity Index, ($SI_{ \delta }$), after normalization. Using shape functions defined as $\sigma \equiv \rho / N$ ($N$ being the number of electrons),  the QSI for density functions simplifies to a QSI for shape functions. This yields the important result that by investigating the similarity between two systems, we are in fact comparing their shape functions. This motivates the investigation of the shape based QSM, found by substitution of $\rho$ by $\sigma$ in expression \ref{ZABdef}.

For the construction of a new QSM, we considered the introduction of  concepts from Information Theory \cite{kulllieb}, which has recently been of increasing relevance to quantum chemical research in general \cite{naleparr} and to the investigation of the electron densities in position and momentum space in particular. (A thorough discussion can be found in the pioneering work of Gadre and Sears \cite{GadreSears}.) 
In our previous work we reported on the information entropy of atomic  
density  and shape functions  respectively defined as

\begin{equation}
	\Delta S_{a} ^{\rho}  \equiv  \int \rho_{a} (\textbf{r}) \log \frac 
{\rho_{a}(\textbf{r})}{\frac{N_{a}}{N_{0}}\rho_{0}(\textbf{r})}d 
\textbf{r}  \label{expr3}
\end{equation}

\begin{equation}
	\Delta S_{a} ^{\sigma}  \equiv \frac{\Delta S_{a} ^{\rho} }{N_a}  =   
\int \sigma_{a} (\textbf{r}) \log \frac{\sigma_{a}(\textbf{r})} 
{\sigma_{0}(\textbf{r})}d\textbf{r} \label{tien}
\end{equation}
with $\rho_{0}(\textbf{r})$ the density of the prior or reference  
atom. As motivated in \cite{CPL1} we set the density function of the  
prior equal to the density of the noble gas preceding the atom under  
investigation in the periodic table, scaled by the factor $\frac{N_ 
{a}}{N_{0}}$, where $N_a$ and $N_0$ are the number of electrons,  
respectively of atom $a$ and its reference.
In this way the prior density $\rho_0(\textbf{r})$ and the density of  
atom $a$, $\rho_a(\textbf{r})$, yield the same number of electrons upon  
integration.
It was shown \cite{CPL1} that these quantities reflect the periodic  
evolution of chemical properties in the Periodic Table and that  
Kullback's interpretation can be formulated in terms of chemical  
information stored in the density functions when we make this  
particular choice for the prior densities.

Following the conclusions in \cite{CPL1}, one can see that it would be interesting to compare the Information Entropy, evaluated locally, $\Delta S ^{\rho}_a(\textbf{r}) \equiv \rho_{a} (\textbf{r}) \log \frac{\rho_{a}(\textbf{r})}{\frac{N_{a}}{N_{0}}\rho_{0}(\textbf{r})}$, of two atoms by use of a QSM.  To that purpose the integrand in expression \ref{expr3} is considered as a function, which gives  the Information Entropy locally at a given point $\textbf{r}$.
The construction of the corresponding QSM becomes straightforward by considering the overlap integral 
%\ref{entropyQSI} 
(with Dirac $\delta$ as separation operator) of the local Information Entropies of two atoms $a$ and $b$ 
\begin{eqnarray} %\label{similariteitsdefinitie}
	Z_{ab}(\delta) & = & \int  \rho_{a} (\textbf{r}) \log \frac{\rho_{a}(\textbf{r})}{\frac{N_{a}}{N_{0}}\rho_{0}(\textbf{r})} \rho_{b} (\textbf{r}) \log \frac{\rho_{b}(\textbf{r})}{\frac{N_{b}}{N_{0'}} \rho_{0'}(\textbf{r})} \; d \mathbf{r} \equiv \int \Delta S_{a} ^\rho (\textbf{r}) \Delta S_b^\rho(\textbf{r}) d \mathbf{r}. \label{entropyQSI} 
\end{eqnarray}
A QSI can be defined by normalizing the QSM as before, via expression \ref{SIdef}. 
The QSM and the normalized QSI give a quantitative way of studying the resemblance in the information carried by the valence electrons of two atoms. 
Expression \ref{entropyQSI} can be rewritten in the form \ref{SIdef} by identification of the operator $\Omega  [  \rho_a (\mathbf{r_1}), \rho_b (\mathbf{r_2}) ; \rho_0 (\mathbf{r_1}), \rho_0' (\mathbf{r_2} ) ]  = \ln \frac{\rho_a(\mathbf{r_1})}{\frac{N_a}{N_0} \rho_0(\mathbf{r_1})} \ln \frac{\rho_b(\mathbf{r_2})}{\frac{N_b}{N_0'} \rho_0'(\mathbf{r_2})} \delta(\mathbf{r_1} - \mathbf{r_2}) $, where we explicitly write the functional dependence on $\rho_a(\mathbf{r_1})$ and $ \rho_b(\mathbf{r_2})$ and the parametrical dependence on $\rho_0(\mathbf{r_1})$ and $\rho_0'(\mathbf{r_2})$. 

The obtained QSI trivially simplifies to a shape based expression
\begin{equation}
SI_{(\delta)}=\frac{\int \Delta S_a ^\sigma (\textbf{r}) \Delta S_b ^\sigma (\textbf{r}) d \mathbf{r}}
{\sqrt{ \int \Delta S_a ^\sigma (\textbf{r}) \Delta S_a ^\sigma (\textbf{r}) d \mathbf{r}}
\sqrt{\int \Delta S_b^\sigma(\textbf{r}) \Delta S_b^\sigma(\textbf{r})d \mathbf{r}}} ,
\end{equation}
where a shorthand notation is used by omitting the explicit dependency of $\textbf{r}$.
The simplification can be generalized from the local information distance operator $\Delta S ^\rho(\textbf{r})$ to any operator $F^\rho(\textbf{r})$, 
which is linear in $\rho$ (thus satisfying $F^\rho(\textbf{r}) = N F^\sigma {(\textbf{r})}$), as follows~:
\begin{eqnarray}
& &\frac{\int F^\rho_a (\textbf{r}) F^\rho_b (\textbf{r})d \mathbf{r}}{\sqrt{\int F^\rho_a (\textbf{r}) \, F^\rho_a (\textbf{r}) d \mathbf{r}} \; \sqrt{\int F^\rho_b (\textbf{r}) \, F^\rho_b (\textbf{r}) d \mathbf{r} }} = \frac{\int F^\sigma_a (\textbf{r}) F^\sigma_b (\textbf{r}) d \mathbf{r}}{\sqrt{\int F^\sigma_a (\textbf{r}) \, F^\sigma_a (\textbf{r}) d \mathbf{r}} \; \sqrt{\int F^\sigma_b (\textbf{r}) \, F^\sigma_b (\textbf{r}) d \mathbf{r}}}  . \label{QSIsigma}  
\end{eqnarray}

In agreement with the fact that the shape function completely determines the properties of a system, as discussed in \cite{BullCarb2004}, the relevance of the QSI as a tool to compare physical properties of atomic electron density functions is confirmed. This characteristic distinguishes the QSI above, together with the Carb\'{o} QSI from other similarity measures (e.g. Euclidian distance \cite{Cioslowski}, Tanimoto \cite{Taminoto} and Hodgkin-Richards \cite{Hodgkins}). 

\subsection{Atomic electron density functions} \label{atomdensity}

\subsubsection{Non-relativistic atomic electron density functions} \label{atomdensity_NR}

The atomic electron density functions were evaluated from non-relativistic numerical Hartree-Fock wave functions optimized on the  $LS$-term ground state of neutral atoms (nuclear charge $3 \leq Z \leq 103 $), as specified in the Table of electron configuration and term value given by Bransden and Joachain or the NIST website~\cite{BraJoa,NISTweb}. 
The extension  of the original Froese-Fischer's code by Gaigalas \cite{gaedimi} allows the calculation of 
term-dependent Hartree-Fock orbitals for any single open subshell case. However, for the two-open subshells cases, this version is limited to 
$(ns)(n'l)^N$~($l = 0, 1, 2, 3, \ldots)$, $(np)^N(n'l)$~($l = 0, 1, 2, 3, \ldots)$ and $(nf)(n'd)$ configurations. This computer code then covers all the ground levels of the periodic table, except atoms with ground configuration $f^{N \geq 2} d $. For these specific systems, we used the  ``MCHF atomic-structure package ATSP2K''~\cite{ATSP-2K}, relying on the combination of the second-quantization approach in the coupled tensorial form, the generalized graphical technique and angular momentum theory in orbital, spin and quasispin spaces and on the use of reduced coefficients of fractional parentage~\cite{nref4}.

In the non-relativistic Hartree-Fock approximation,  in its single-configuration version, the atomic wave function is limited to one configuration state function (CSF)~\cite{zeventien}
\[
\vert \alpha \pi L S M_L M_S \rangle , \]
  simultaneous eigenfunction of the inversion operator, the orbital angular momentum {\bf L}$^2$, the spin angular momentum ${\bf S}^2$ and their projections $L_z$ and $S_z$, that can be built from the one-electron spin-orbitals
 \begin{equation}
 \psi_i(r) = R_{nl}(r) Y^m_l (\theta, \phi) \chi_{1/2,m_s}(\sigma)\equiv r P_{nl}(r) Y^m_l (\theta, \phi) \chi_{1/2,m_s}(\sigma).
 \end{equation} 
$\alpha$ denotes all the information needed to specify unambiguously the term considered (configuration and coupling tree). 
% ie. the configuration, the subshells coupling tree with, for each subshell, the spin and orbital quantum numbers, 
% the seniority, and, for the $f$-subshell, an extra group label if necessary.
The optimized one-electron numerical radial functions $\{ P_{nl}(r) \}$ are used to determine the corresponding $LS$-dependent electron density function from the following expression
\begin{eqnarray}
	\rho(r) & = &  \frac{1}{4 \pi} \;  
	\sum_{nl} \frac{P_{nl}^{2}(r)}{r^{2}} \; q_{nl} \label{dichtheid} ,
\end{eqnarray}
where $q_{nl}$ is the occupation number of the subshell considered. 
In the case of uncompletely filled subshells, spherical averaging over the $(M_L,M_S)$ term components was applied, yielding a spherical electron density function, as elaborated in \cite{CPL1}. In an $LS$-dependent Hartree Fock scheme, the radial wave functions are allowed to vary, for a given electronic configuration,  from one term to another. 
Eq.~\ref{dichtheid} should then strictly be read as 
\begin{eqnarray}
\rho(r) \equiv \rho_{\alpha LS} (r) \label{eq4}
\end{eqnarray}

An easy way of testing the calculated density function $\rho(r)$ is to check that its integration yields the total number of electrons
\begin{eqnarray}
    4 \pi \int_0^\infty \rho(r) r^2 dr = \sum_{nl} q_{nl} = N ,
\end{eqnarray}
as expected from the normalization constraint of the Hartree-Fock numerical one-electron radial wave functions.

\subsubsection{Relativistic atomic electron density functions} \label{atomdensity_R}

For the purpose of quantifying the relativistic effects on the electron density functions, we evaluate the similarity of Hartree-Fock and Dirac-Fock density functions using a point nucleus aproximation. In the relativistic scheme, the atomic wave function is, in the most general case, a combination of configuration state functions (CSF's)
\begin{equation}
\vert  \pi J M_J \rangle = \sum_{\nu} c_\nu\vert \nu \pi J M_J \rangle 
\end{equation}
eigenfunction of the inversion operator, the total angular momentum {\bf J}$^2$ and its projection $J_z$.  
$\nu$ denotes all the necessary information for specifying the relativistic configuration completely. The CSF are built on the one-electron Dirac four-spinor
\begin{equation}
\psi_i(r) = \frac{1}{r}
\left( 
\begin{array}{c}
P_i(r) \chi_{\kappa_i}^{\mu_i} (\Omega) \\ 
i Q_i(r) \chi_{- \kappa_i}^{\mu_i} (\Omega) \end{array} \right)
\end{equation}
where $\chi_{\kappa_i}^{\mu_i} (\Omega)$ is a two-dimensional vector harmonic. It has the property that $K \psi_i(r) = \kappa \psi_i(r)$ where 
$K = \beta ( \mbox{\boldmath $ \sigma $} \cdot \bf{L} + 1).$

 The large $\{ P(r) \}$ and small $\{ Q(r) \}$ components are solutions of a set of coupled integro-differential equations \cite{nref5}. The  mixing coefficients $\{c_{\nu} \}$ are obtained by diagonalizing the matrix of the no-pair Hamiltonian containing the magnetic and retardation terms~\cite{nref6}. The two coupled variational problems are solved iteratively. For a complete discussion on relativistic atomic structure we refer to \cite{IPGrant96}. The present calculations have been performed using the MDF/GME program of Desclaux and Indelicato \cite{MDFGME} including both the magnetic and retardation part of the Breit interaction in the self-consistent process, but not the vacuum polarization.
 
It is to be noted that the relativistic scheme rapidly becomes more complicated than the corresponding non-relativistic one.
For example, if the ground term of Carbon atom is described, in the non-relativistic one-configuration Hartree-Fock approximation, by a single CSF 
$\vert 1s^2 2s^2 2p^2 \; ^3P \rangle  $, the relativistic equivalent implies the specification of the $J$-value. For $J=0$ corresponding to the ground level of Carbon,  the following  two-configuration description becomes necessary
\[
\vert `` 1s^2 2s^2 2p^2 " (J=0) \rangle = c_1 \vert 1s^2 2s^2 (2p^\ast)^2 (J=0) \rangle +  c_2 \vert 1s^2 2s^2 2p^2 (J=0) \rangle ,
\]
implicitly taking into account the relativistic mixing of the two $LS$-terms ($\;^1S$ and $\;^3P$) arising from the $2p^2$ configuration and belonging to the $J=0$ subspace.  $p^\ast$ and $p$ correspond to the $j$-values, $j=1/2$ ($\kappa = +1$) and $j=3/2$ ($\kappa = -2$), respectively.

By averaging the sublevel densities, 
\begin{equation}
\rho ({\bf r}) 
= \frac{1}{(2J+1)} \sum_{M_J=-J}^{+J}
\; \rho^{J M_J} ({\bf r})
\end{equation}
the total electron density becomes spherical for any open-shell system, as found in the non-relativistic scheme~\cite{CPL1}, and can be calculated from
\begin{eqnarray}
	\rho (r) & = &  \frac{1}{4 \pi} \;  
	\sum_{n \kappa} \frac{P_{n \kappa}^{2}(r) + Q_{n \kappa}^{2}(r) }{r^2} \; q_{n\kappa} \label{rel_dichtheid} ,
\end{eqnarray}
where $q_{n\kappa}$ is the occupation number of the relativistic subshell $(n \kappa)$.

\section{Information Theory QSM and QSI for atomic density functions}

\begin{figure}[!h]
    \begin{center}
      \resizebox{ 150mm}{!}{\includegraphics{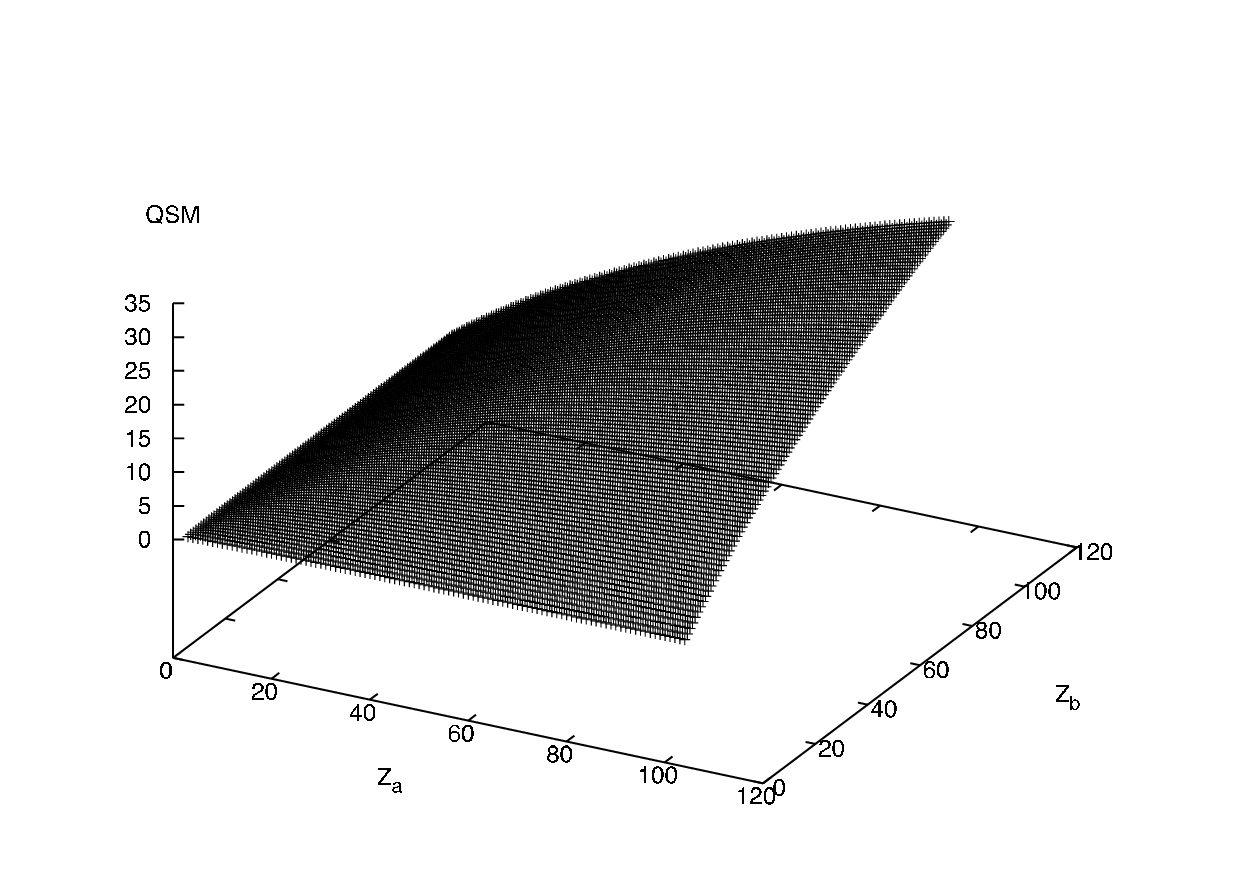}}
      \caption{Overlap integral of the non-relativistic Hartree-Fock shape functions (the QSM appearing in the definition of the Carb\'{o} QSI) evaluated for all pairs of atoms. A monotonic trend of increasing QSM for heavier atoms is revealed. The vertical axis corresponds to the QSM of the atoms with nuclear charges $Z_a$ and $Z_b$ given by the axes in the plane.}
      \label{carbom}
    \end{center}
  \end{figure}

\begin{figure}[!h]
    \begin{center}
      \resizebox{ 150mm}{!}{\includegraphics{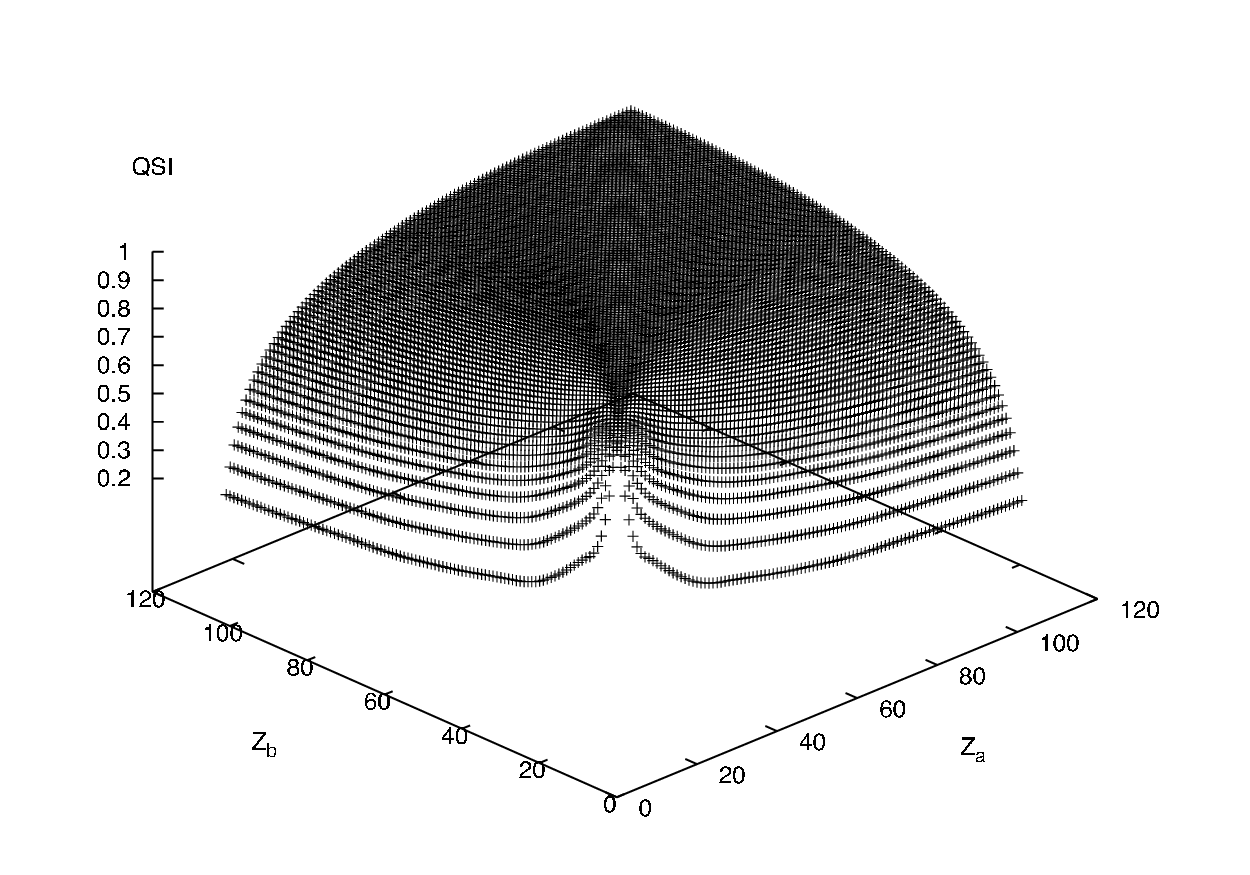}}
      \caption{The Carb\'{o} QSI evaluated for all pairs of atomic non-relativistic Hartree-Fock shape functions, revealing a nearest neighbour effect \cite{CPL1}. The vertical axis corresponds to the QSI of the atoms with nuclear charges $Z_a$ and $Z_b$, indicated by the axes in the plane.}
      \label{carbo}
    \end{center}
  \end{figure}

\begin{figure}[!h]
    \begin{center}
      \resizebox{ 150mm}{!}{\includegraphics{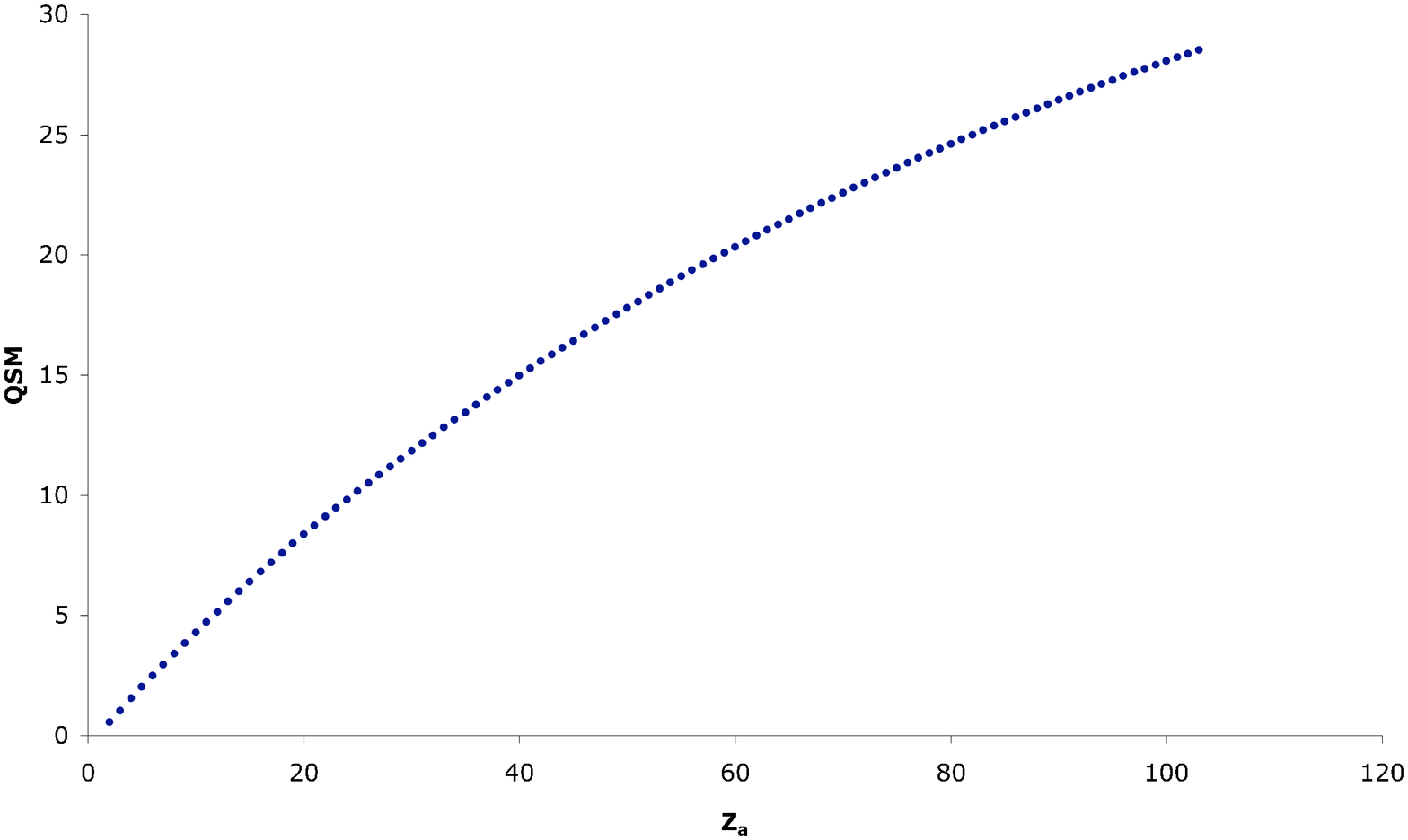}}
      \caption{The cross section of the graph in figure \ref{carbom} for $Z_b = 82$. The vertical axis gives the QSM value for $Pb$ with the atom with nuclear charge $Z_a$.}
      \label{carbom_82}
    \end{center}
  \end{figure}

In this section the results of the QSM and QSI, evaluated for shape functions of all pairs of atoms in the periodic table are discussed.
To facilitate the interpretation of the results of the Information Theory based QSM and QSI  a graphical representation of the Carb\'{o} QSM (figure \ref{carbom})
 and QSI (figure \ref{carbo}), already mentioned in \cite{CPL1}, is given. The results of all the possible pairs of atomic shape functions are given in
 a $3$ dimensional graph, where the vertical axis indicates the QSM or QSI value of the atoms with nuclear charges $Z_a$ and $Z_b$. These figures show
 that the general trend for any fixed $Z_a$ is similar for all atoms: all the cross-sections of the $3$ dimensional graphs show the same evolution.
In figure \ref{carbom_82} we show the cross-section of the results of the QSM between Pb ($Z=82$) and all other atoms. The general trend of the overlap QSM of the density functions increases monotonically with increasing volume of the atoms, as pointed out in \cite{twaalf}.
 
\begin{figure}[!h]
    \begin{center}
      \resizebox{ 150mm}{!}{\includegraphics{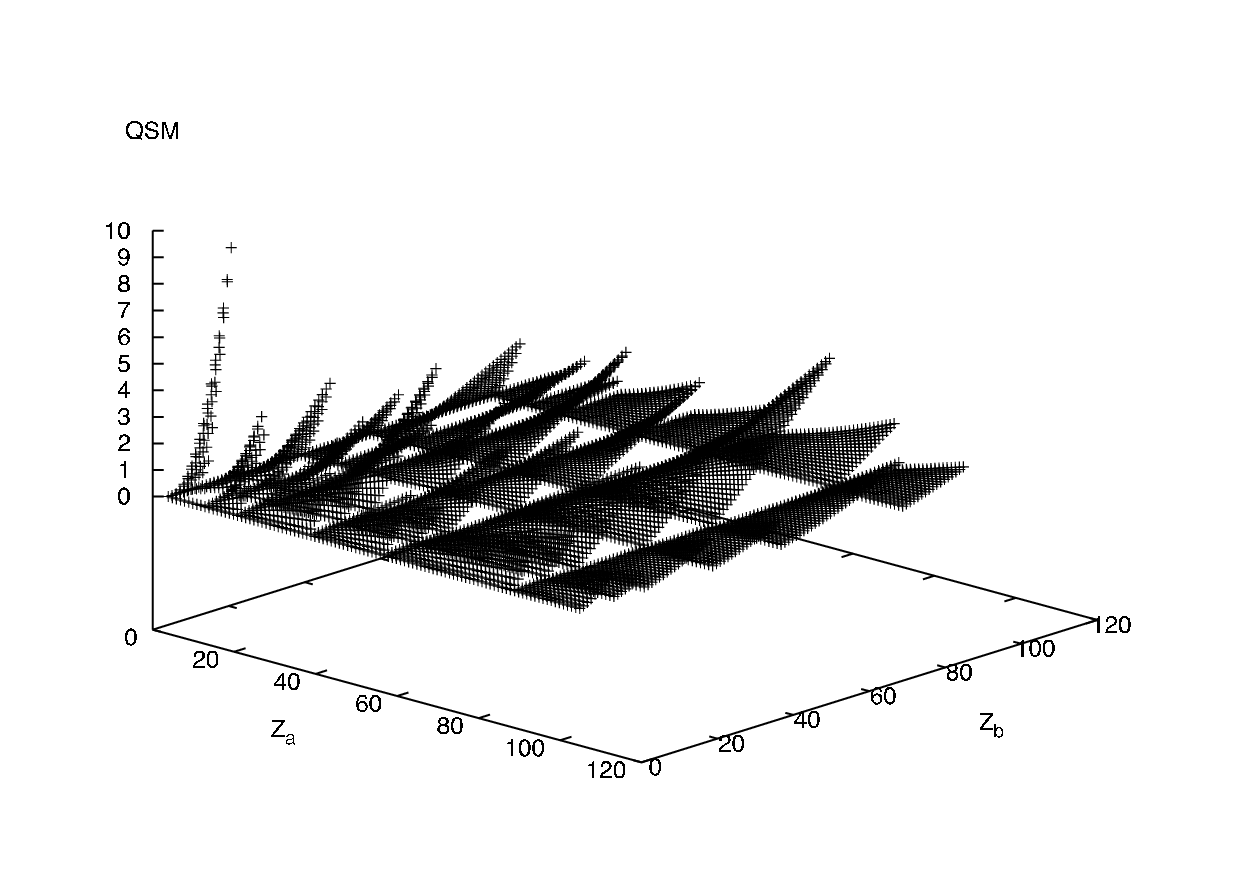}}
      \caption{ Information Entropy based QSM for pairs of atoms in the periodic table, with the noble gas of the previous row as prior for each given atom. A clear periodic character can be distinguished. A non-relativistic Hartree-Fock approach was used.}
      \label{siklm}
    \end{center}
  \end{figure}

\begin{figure}[!h]
    \begin{center}
      \resizebox{ 150mm}{!}{\includegraphics{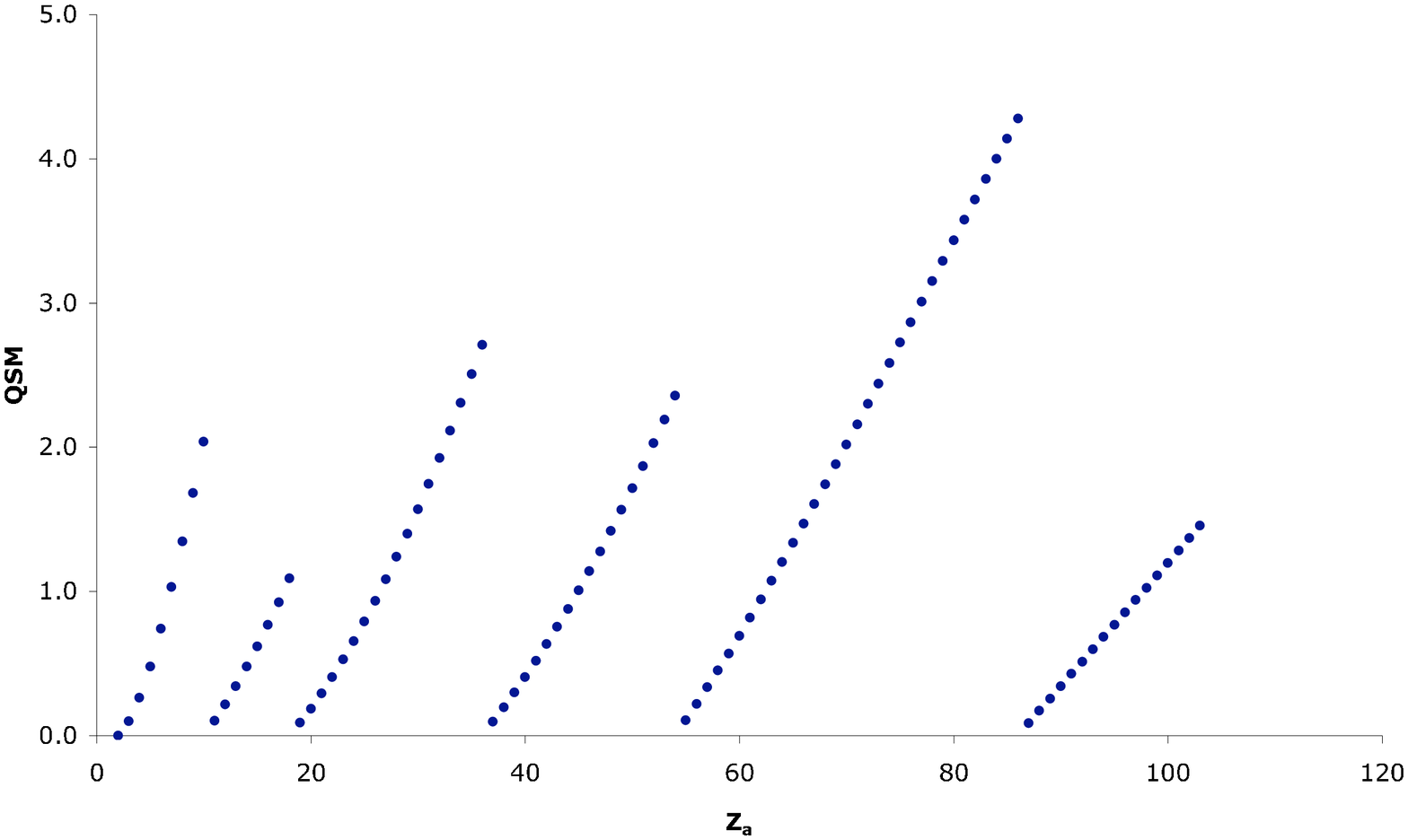}}
      \caption{ The cross section of the graph in figure \ref{siklm} for $Z_b = 82$. The periodic character is regained.}
      \label{siklm_82}
    \end{center}
  \end{figure}

\begin{figure}[!h]
    \begin{center}
      \resizebox{ 150mm}{!}{\includegraphics{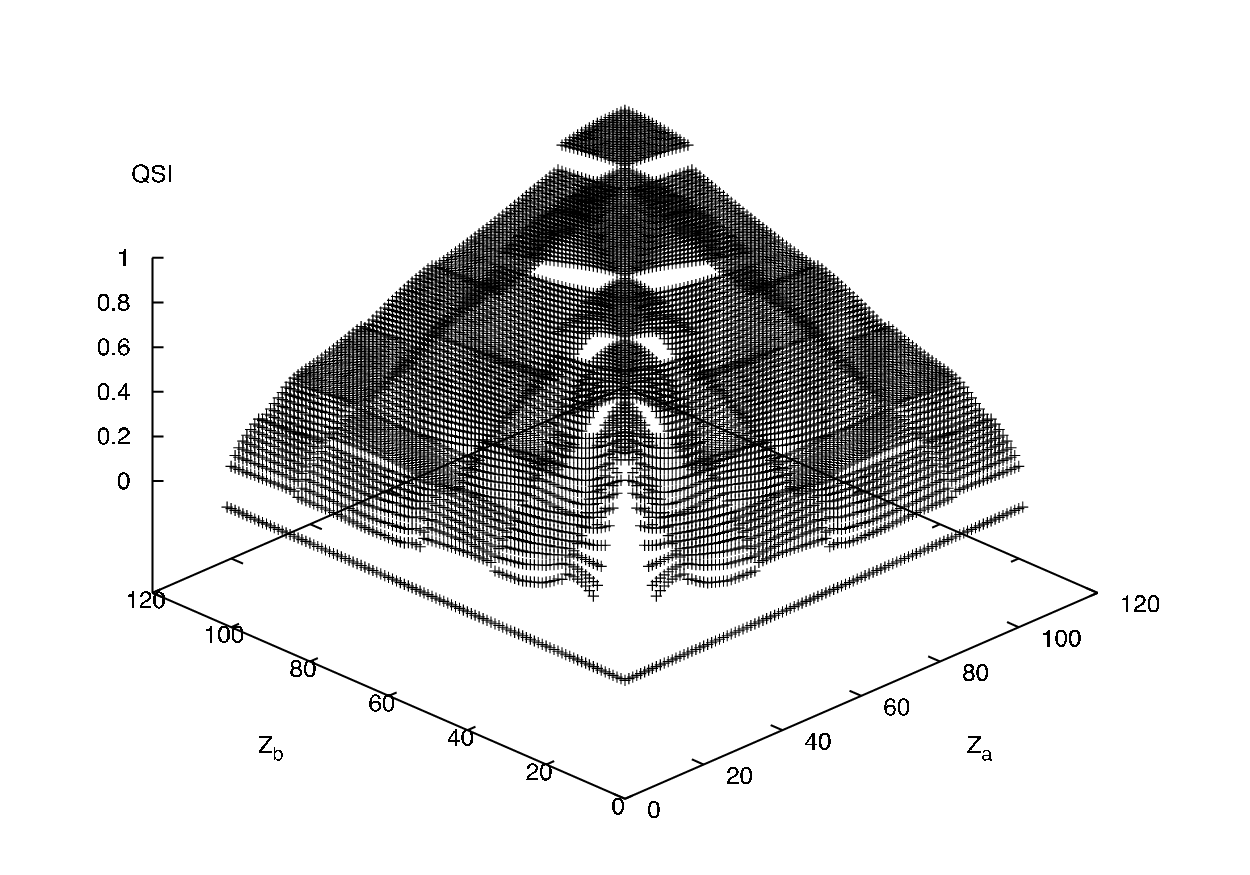}}
      \caption{ Information Entropy based QSI for pairs of atoms in the periodic table, with the noble gas of the previous row as prior for each given atom. The vertical axis corresponds to the QSI of the atoms with nuclear charges $Z_a$ and $Z_b$ given by the axes in the plane.}
      \label{sikl}
    \end{center}
  \end{figure}

\begin{figure}[!h]
    \begin{center}
      \resizebox{ 150mm}{!}{\includegraphics{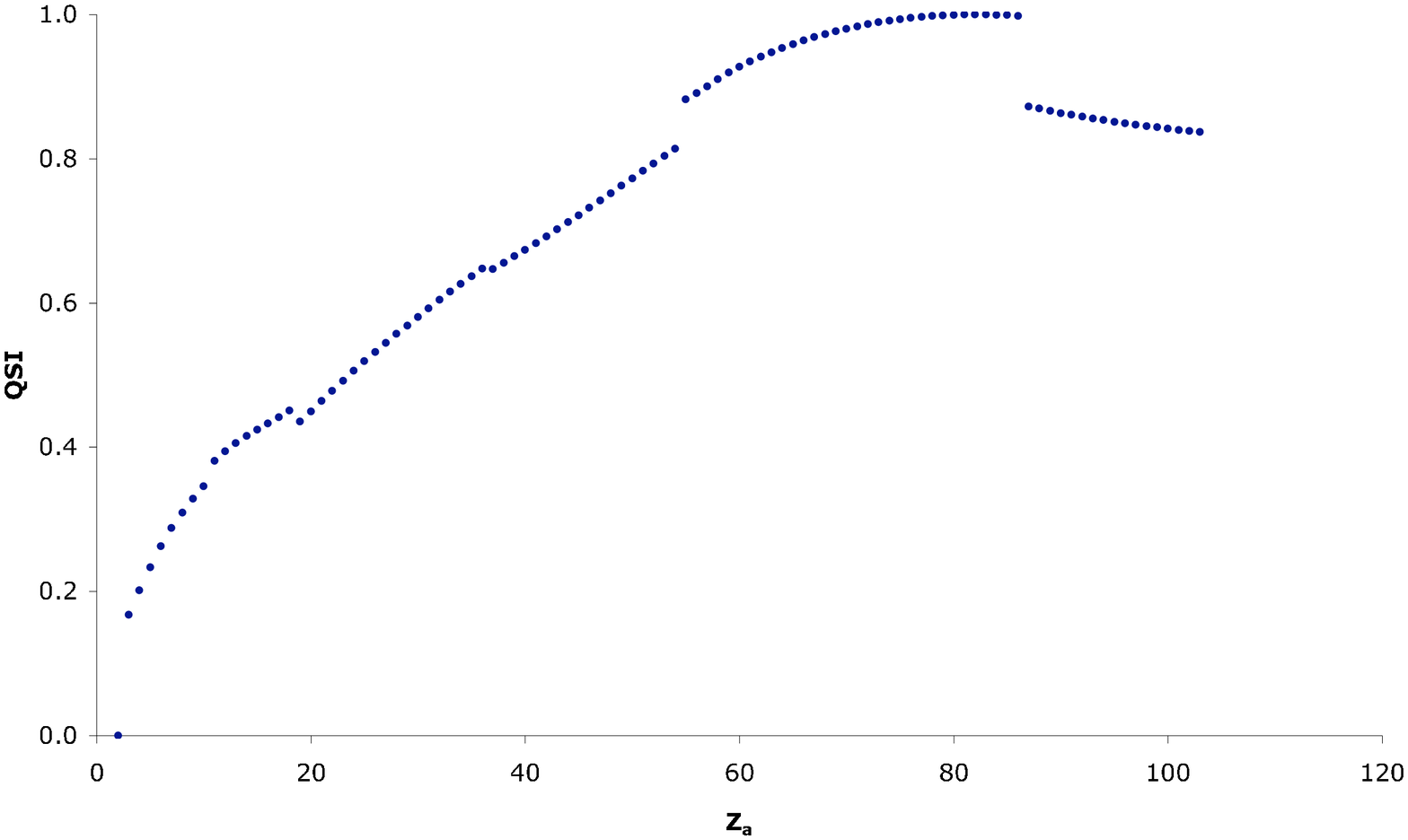}}
      \caption{ A cross section for  $Z_b = 82$ of the graph of Information entropy QSI in figure \ref{sikl}. The change of reference atom is still visible, but the periodicity is not regained.}
      \label{sikl_82}
    \end{center}
  \end{figure}

The evaluation of the Information Theory based QSM (figure \ref{siklm}) was found to be positive for all investigated atoms  reveals a picture corresponding to the periodicity of Mendeleev's Table, which can be distinguished by looking at the cross-section for Pb in figure \ref{siklm_82}. The results correspond to the evolution of chemical properties first of all in the sense that for each period the QSM increases gradually from the first column to the last. Ionization energy and Hardness are properties which reveal a similar evolution throughout \cite{LiuDeProft1997}. Secondly in the sense that neighboring atoms with large nuclear charge differ less than neighboring light atoms, e.g. the difference between the QSM values of two atoms in the first period is large in comparison to the difference in QSM between two neighboring Lanthanides.
Considering all the cross-sections of Figure \ref{siklm} reveals that the periodicity is regained throughout by the choice of the reference atoms, as it yields low QSM values for atoms similar to the chosen prior. 
One notes however that the QSM does not reveal results, which reach maxima when a given atom is compared with another atom of the same group.
Moving to the QSI, the periodicity of the QSM is lost due to the normalization (figures \ref{sikl} and \ref{sikl_82}). In figure \ref{sikl_82} the change of prior is still visible due to the gaps (discontinuities) at the positions where the prior changes, but the normalization blends out the clear periodic evolution of the QSM in graph \ref{siklm_82}.
This leads to the conclusion that the normalization, which yielded the nearest neighbor effect for the Carb\'{o} QSI in figure \ref{carbo}, can overwhelm the characteristics of a QSM.


\begin{figure}[!h]
    \begin{center}
      \resizebox{ 150mm}{!}{\includegraphics{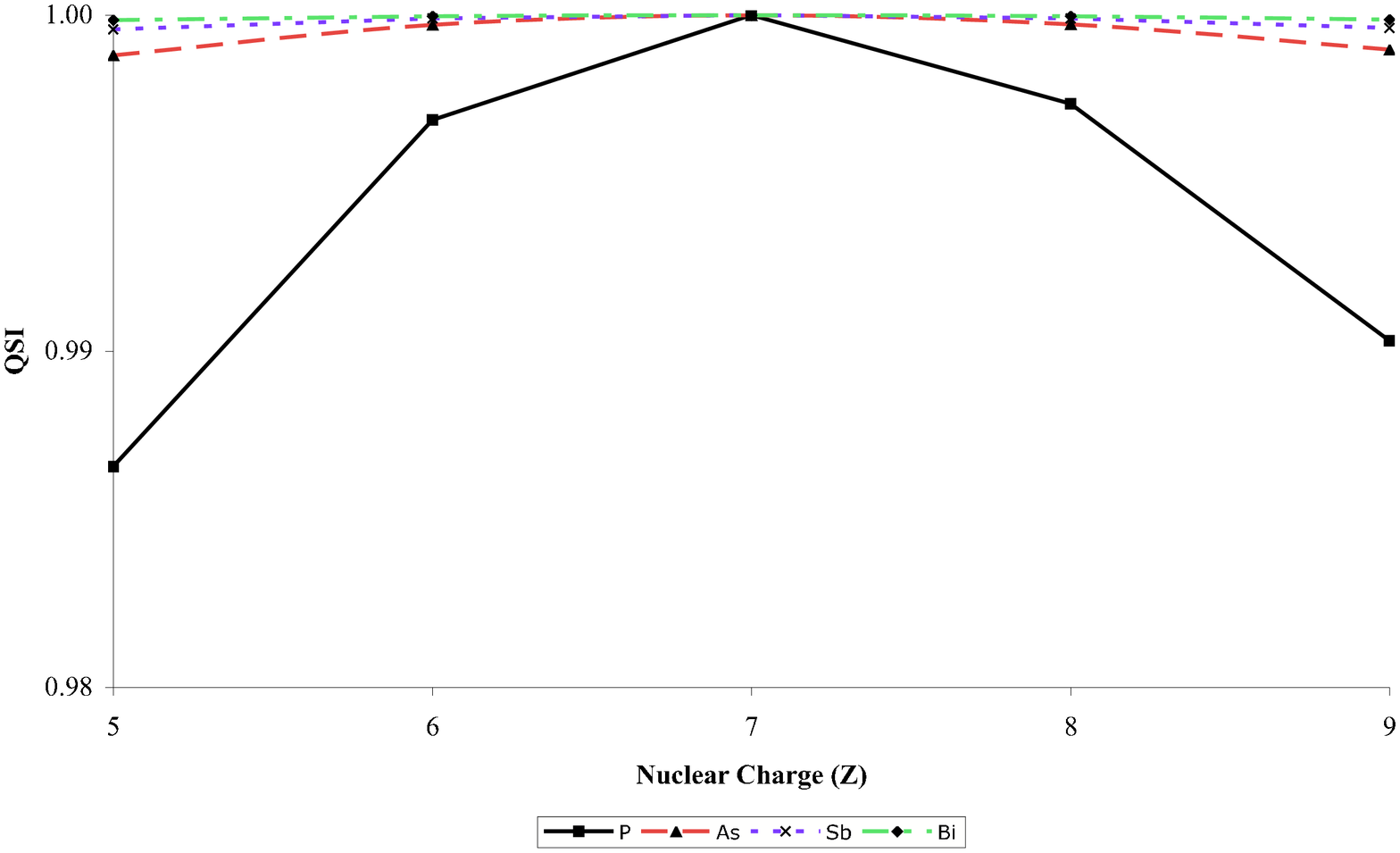}}
      \caption{ Results of the information theory based QSI with the atom on top of the column as prior. The symbol in the legend indicates the period of the investigated atom and the nuclear charge $Z$-axis indicates the column of the investigated atom. (For example $Ga$ can be found as a square $Z=5$). }
      \label{plotcol}
    \end{center}
  \end{figure}

Changing the point of view, we can opt to investigate which atom of a given period of the table belongs to a certain column and in which way the atoms should be ordered within the period. This can be done by investigating the QSI with the top atoms of each column as prior. Formulated in terms of Kullback Liebler information discrimination the following is evaluated.
For instance, when we want to investigate the distance of the atoms $Al$, $Si$, $S$ and $Cl$ from the $N$-column (group Va),
 we consider the information theory based QSI in expression \ref{entropyQSI}, where the reference densities $\rho_0$ and $\rho_{0'}$ are set to $\rho_N$, $\rho_A$
 to $\rho_{Al}$, $\rho_{Si}$,  $\rho_{P}$, etc. respectively and $\rho_B$ to $\rho_{P}$, i.e. we compare the information contained in the shape function of $N$ to determine
 that of $P$, with its information on the shape function of $Al$, $Si$, $S$, $Cl$. The data in table \ref{tablecol} reveal a $1.$ for the element $P$
 (by construction) with values continuously decreasing from unity for the elements to the left and to the right of the $N$-column.
 This pattern is followed for the periods $3$ up to $6$, taking $As$, $Sb$ and $Bi$ as reference, with decreasing difference along a 
given period (see Figure \ref{plotcol}) in accordance with the results above. Note that the difference from $1.$ remains small, due to the
 effect of the renormalization used to obtain the QSI.
\begin{table}[!h]
\caption{ Numerical results of the QSI with prior atoms set to the elements on top of the columns. The information present in the shape function of $N$ to obtain information on that of $A$ is compared with the information present in the shape function of $N$ to obtain information about $N$-group atom of the corresponding period. }
\begin{center}
\begin{tabular}{|c|c|c|c|c|}
\hline
Al: 0.98656 &Si: 0.99688 &P: 1. & S: 0.99735 & Cl: 0.99031 \\
Ga: 0.99880 &Ge:  0.99971 & As: 1. & Se: 0.99973 & Br: 0.99897\\
In: 0.99957 & Sn: 0.99989 &Sb: 1. & Te: 0.99990 & I:  0.99961\\
Tl: 0.99986 &Pb: 0.99996 &Bi: 1. & Po: 0.99996 & At: 0.99987 \\
\hline

\end{tabular}
\end{center}
\label{tablecol}
\end{table}

%The results of the $B$ (indicated above $Z=5$) column reveal that $Al$ differs more from $Si$ than $Ga$ from $Ge$ etc. This indicates that the difference between neighboring atoms increases with the nuclear charge $Z$.

%Alternatively formulated one obtains: to investigate if atom $B$ belongs to the same column as atom $A$ one can use the QSI defined directly above, by setting the prior to the density function of the atom on top of the column. If they are in the same column, the QSI yield 1., by construction of the QSI and yields numbers decreasing monotonically with the distance between the columns. This is an encouraging result, nevertheless it should be noted that a lot of parameters have to be tuned to get this result.
 
\section{Investigation of relativistic effects} \label{relat}

\begin{figure}[!h]
    \begin{center}
      \resizebox{ 150mm}{!}{\includegraphics{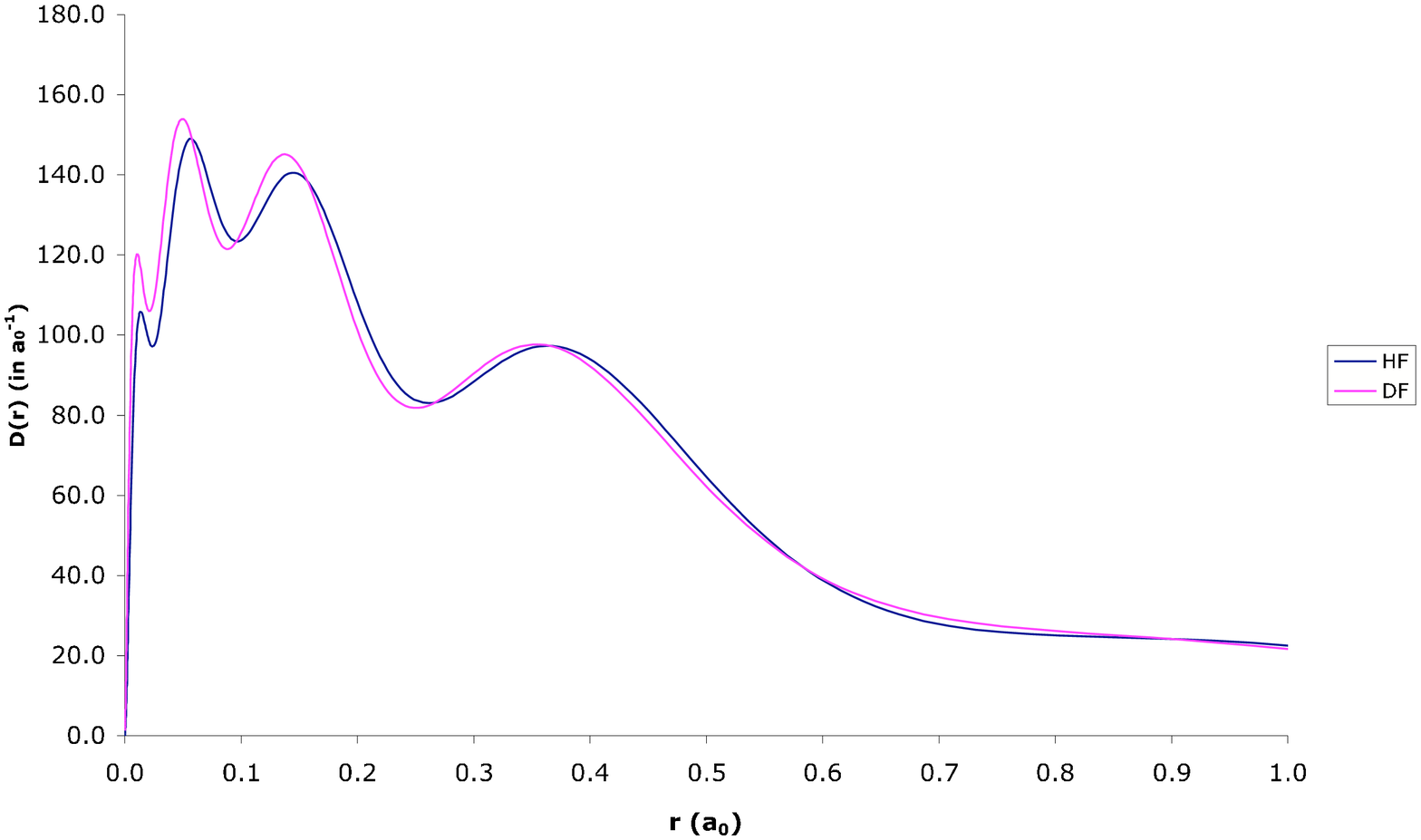}}
      \caption{DF and HF density distributions $D(r) = 4 \pi r^2 \rho(r) $ for the neutral Pb atom ($Z=82$). The contraction of the first shells is clearly visible.}
      \label{Pbraddist}
    \end{center}
  \end{figure}

In this section we discuss the relativistic effects on atomic electron density functions. 
We first illustrate the difference of the radial density functions $D(r)$ defined  as~\cite{BraJoa}, 
\begin{equation}
D(r) \equiv 4 \pi r^2 \rho(r)  ,  \label{dist} 
\end{equation}
calculated in the Hartree-Fock (HF) and Dirac-Fock (DF) approximations for the ground state $6p^2 \; ^3P_0$ of Pb~I ($Z=82$) according to equations~\ref{dichtheid} and \ref{rel_dichtheid}, respectively.
These are plotted in figure~\ref{Pbraddist}, which shows the global relativistic contraction of the shell structure.

\begin{figure}[!h]
    \begin{center}
      \resizebox{ 150mm}{!}{\includegraphics{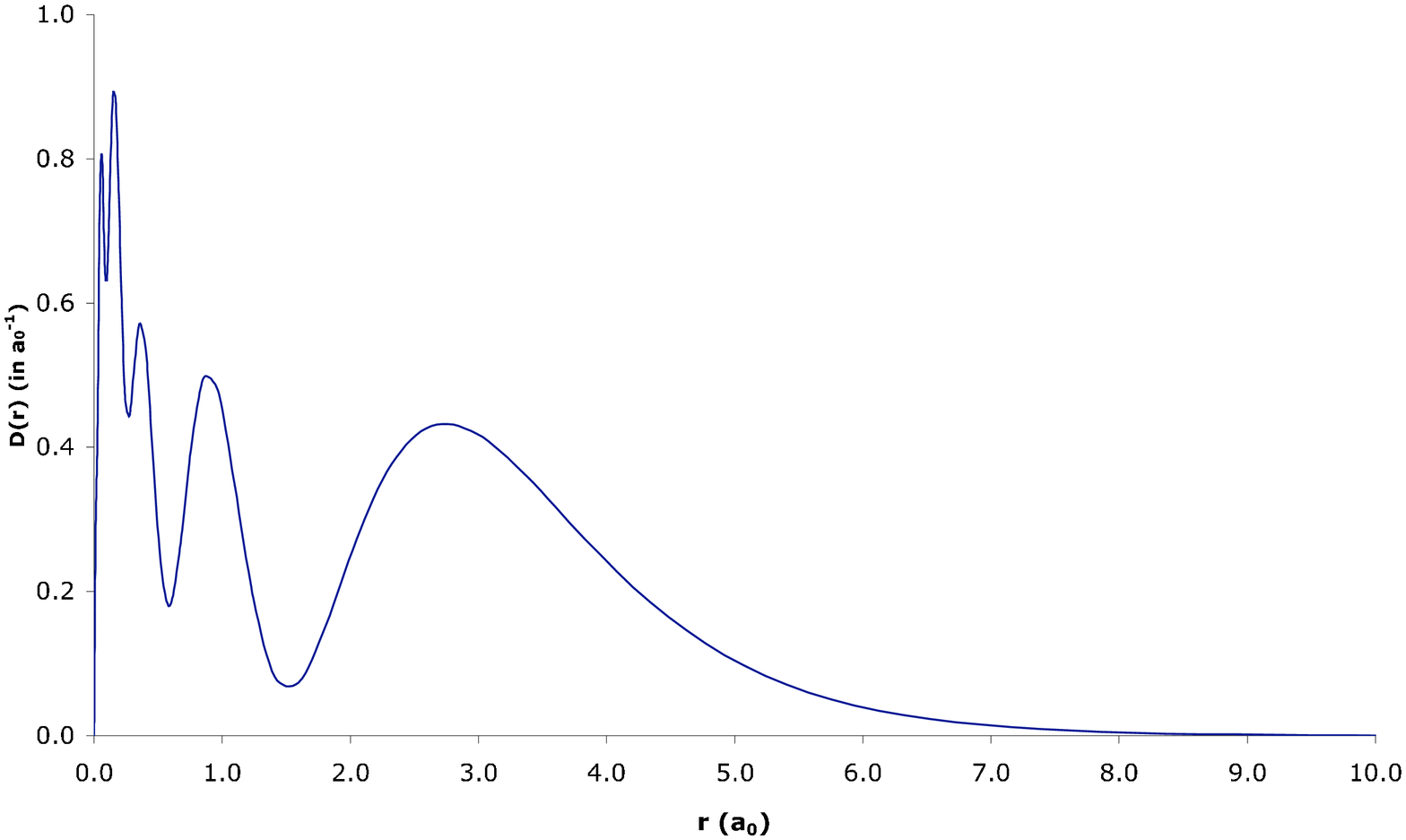}}
      \caption{Accumulated difference between the DF and the HF densities of Pb. }
      \label{Pb_dist}
    \end{center}
  \end{figure}

Another interesting observation can be done from figure~\ref{Pb_dist} displaying, as was done in \cite{Sen} 
for the Rn atom, the accumulated difference between the DF and the HF radial density functions \ref{dist} defined as
\begin{equation}
\Delta D(r) \equiv \int_0^r  \left( D^{\mbox{DF}}(r') - D^{\mbox{HF}}  (r') \right)  dr'
\end{equation}
as a function of $r$, the radial distance to the nucleus.
We see in this way that there is an excess charge, varying between $0$ and $0.9$ due to relativistic effects. One notices that the contraction of the total radial density function reveals a shell structure. 
Since the densities are normalized to the same number of electrons, the accumulated difference converges to $0$ for large values of $r$.

Employing the framework of QSI to compare non-relativistic Hartree-Fock electron density functions $\rho^{HF}(\mathbf{r})$ with relativistic Dirac-Fock electron density functions $\rho^{DF}(\mathbf{r})$ for a given atom, the influence of relativistic effects on the total density functions of atoms can be quantified via the QSI defined below

\begin{eqnarray} 
	& Z_{HF,DF}(\delta)  =  \int  \rho^{HF}(\mathbf{r})  \, \rho^{DF}(\mathbf{r}) \; d \mathbf{r}  \\
	& SI_{ \delta }  =  \frac{Z_{HF,DF}(\delta) } {\sqrt{Z_{HF,HF}(\delta) }\sqrt{Z_{DF,DF}(\delta) }}, \label{QSIhfvsdf}
\end{eqnarray}
where $\delta$ is the Dirac-$\delta$ operator.
 \begin{figure}[!h]
    \begin{center}
      \resizebox{ 150mm}{!}{\includegraphics{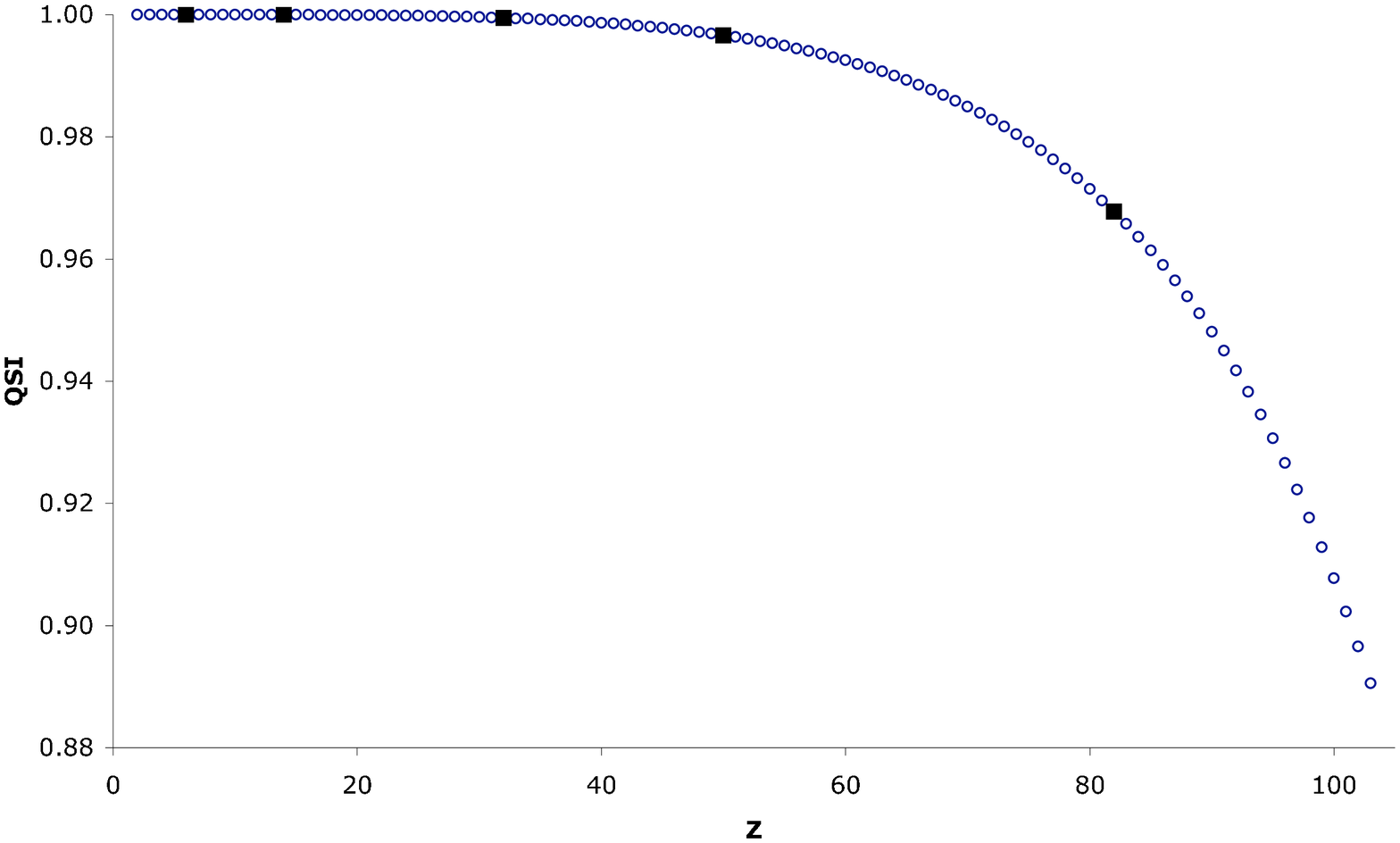}}
      \caption{Similarity of non-relativistic Hartree-Fock with relativistic Dirac-Fock atomic density functions with highlighted results for the CIV group atoms.}
      \label{hfvsdf}
    \end{center}
  \end{figure}

In figure \ref{hfvsdf} we supply the QSI between atomic densities obtained from numerical Hartree-Fock calculation and those obtained from numerical Dirac-Fock calculations, for all atoms of the periodic table. 
The results show practically no relativistic effects on the electron densities for the first periods, the influence becoming comparatively large for heavy atoms. 
To illustrate the evolution through the table the numerical results of the carbon group elements are given in table \ref{cgroup} and highlighted in the graph in figure \ref{hfvsdf}. From the graph it is also noticeable that the relativistic effects rapidly gain importance for atoms heavier than Pb.

\begin{table}[!h]
\caption{ Numerical results of the QSI for the carbon group, highlighted in graph \ref{hfvsdf}.} 
\begin{center}
\begin{tabular}{|c|c|}
\hline
Atom & QSI \\
\hline
C & 0.99999 \\
Si & 0.99996 \\
Ge & 0.99945 \\
Sn & 0.99661 \\
Pb & 0.96776 \\
\hline

\end{tabular}
\end{center}
\label{cgroup}
\end{table}

\begin{figure}[!h]
    \begin{center}
      \resizebox{ 150mm}{!}{\includegraphics{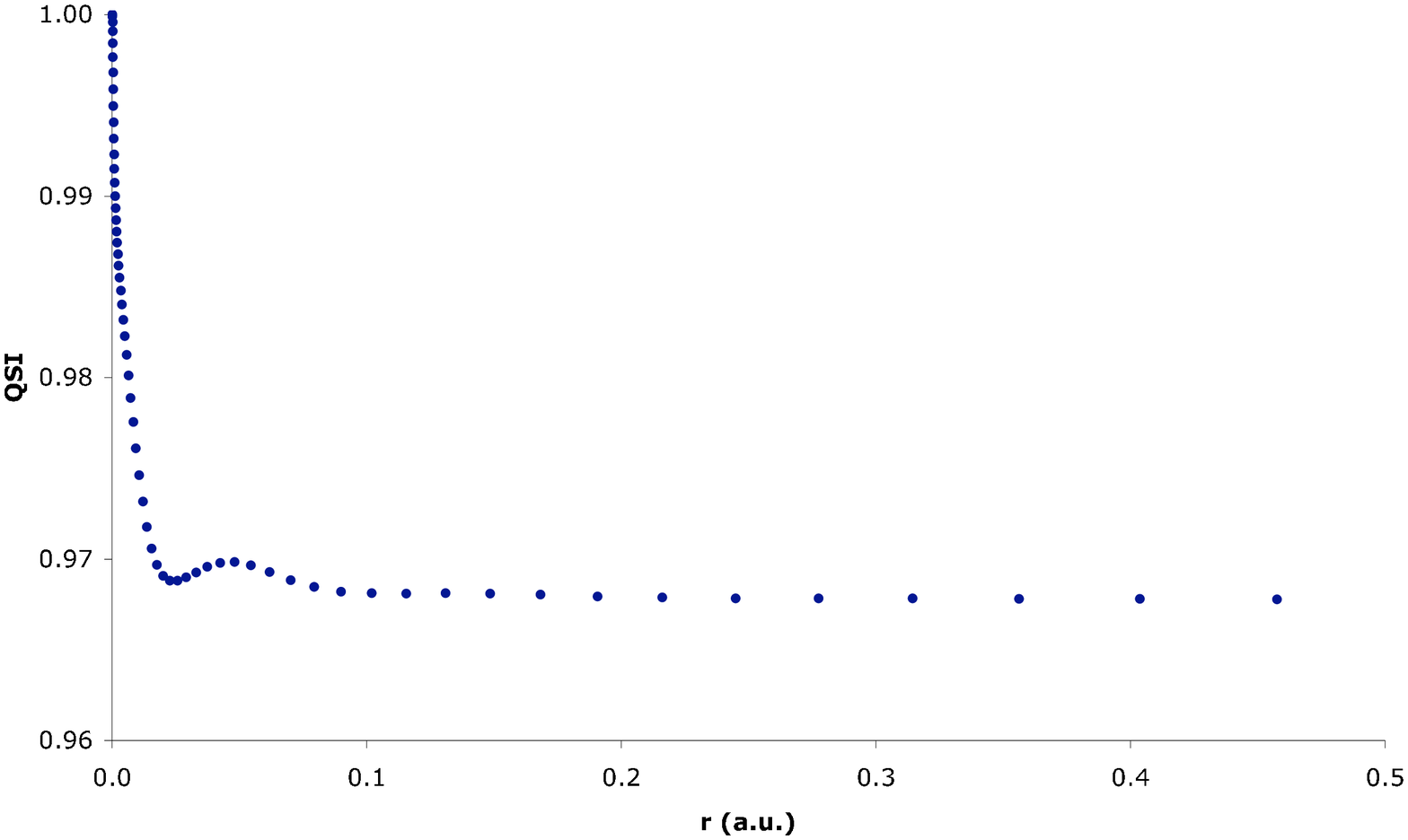}}
      \caption{Convergence of the QSI of HF and DF densities of Pb, as a function of $r$, revealing the large influence of the inner region of the density function. }
      \label{Pb_QSIR}
    \end{center}
  \end{figure}

Investigation of the convergence of the QSI in function of the radius $r$ can shed some light on the importance of core region of the density functions for the QSI. 
In figure \ref{Pb_QSIR} we plot, for the Pb atom, the numerical results of the QSI defined as 
\begin{eqnarray} 
	& Z_{HF,DF}(\delta ;r)  =  \int_0^{r} \int_\Omega  \rho^{HF}(\mathbf{r'})  \, \rho^{DF}(\mathbf{r'}) \; d \Omega dr'  \\
	& SI_{ \delta } (r)  =  \frac{Z_{HF,DF} (\delta ;r) } {\sqrt{Z_{HF,HF} (\delta ;r) }\sqrt{Z_{DF,DF} (\delta ;r) }}  \label{QSIRhfvsdf}
\end{eqnarray}
where the integration over $\Omega$ represents the integration over all angles.
The plot shows a very fast convergence, the total QSI value being reached already for $r=0.2~a_0$ . 
This result demonstrates the dominance of the inner region of the density function for this type of QSI. This picture does not reflect the influence of relativistic effects on the valence electrons, which is visible in the accumulated difference in figure~\ref{Pbraddist}. The rapid convergence of the QSI can be accounted to the fact that the densities, in the overlap integral, themselves are much larger in regions of small radius $r$, whereas the clear influence of relativistic effects in the accumulated difference picture is due to the fact that the difference between Hartree-Fock and Dirac-Fock radial densities remains of the same order of magnitude, converging to zero, as can be seen in figure \ref{Pbraddist}.

\section{Conclusion}

In this communication we report on the development and calculation of a new, Information Theory based, Quantum Similarity Measure (QSM) and its corresponding Quantum Similarity Index (QSI) for atoms, using their density functions and shape functions. 
We show that a QSM constructed with the Kullback Leibler Information Entropy loses its periodic character upon normalization. One might say that the normalisation renders the QSI insensitive to certain characteristics, clearly present in the QSM. 
To regain the periodicity with the information theory based QSM, the choice of the prior for each atom as the density of the noble gas of the previous row, normalized to the same number of electrons in the atom under investigation, is crucial. The results of the QSM are in agreement with chemical intuition in the sense that the difference in QSM of two successive light atoms is large in comparison to the difference in QSM of two successive heavy atoms, meaning that light atoms next to each other in the Table differ more than neighboring heavy atoms. When looking at the results of Lanthanides and Actinides in particular we find high similarities indeed.
This interpretation is not regained by looking at the QSI, with the prior set to the noble gas atoms. It is rewarding that the comparison of information content of the shape function of a given top atom in a column with the atoms of the subsequent period(s) reveals another periodicity pattern. 

The visualization of the influence of relativistic effects on the radial density distribution reveals a shell structured excess charge, corresponding to the contraction of the charge distribution.
The importance of relativistic effects for the electron density functions and shape functions of atoms has been quantified via a study based on QSI. 
A plot of the QSI as a function of the nuclear charge shows that the densities of light atoms are barely influenced by including the relativistic corrections and that the influence of relativistic effects increases monotonically with the nuclear charge of the neutral atom throughout the Periodic Table. 

As suggested by the investigation of: i) the Carb\'{o} QSI, ii) the Information Theory based QSI and iii) relativistic effects via the Carb\'{o} QSI for atoms, the Carb\'{o} QSI reflects the similarity of the core region of the density function, i.e. it fails to reflect the importance of the valence electrons, which is essential from a chemical point of view.
It would be interesting to investigate if the valence region can be given more weight in a similarity study, by introducing an appropriate separation operator in the definition of the QSI.

Although correlation effects are neglected in the present work, a similar QSM/QSI approach can be used for investigating how much the electron densities are affected by correlation, comparing the same atom in the single- and multi-configuration non relativistic Hartree Fock approximations.

\section*{Acknowledgments} 

M.G. thanks the Communauté française of Belgium (Action de 
Recherche Concertée) and the Belgian National Fund for Scientific
Research (FRFC/IISN convention) for their financial support. Laboratoire Kastler Brossel is Unit\'{e} Mixte de Recherche du CNRS
n$^{\circ}$ 8552

% \appendix

% \section{}
% \label{}

\end{document}